# Heterogeneous Processor Pipeline for a Product Cipher Application

Isuru B. Nawinne, Mahanama S. Wickramasinghe, Roshan G. Ragel, *Member, IEEE*, Swarnalatha Radhakrishnan

Department of Computer Engineering, University of Peradeniya, Peradeniya 20400 Sri Lanka

*Abstract*-Processing data received as a stream is a task commonly performed by modern embedded devices, in a wide range of applications such as multimedia (encoding/decoding/ playing media), networking (switching and routing), digital security, scientific data processing, etc. Such processing normally tends to be calculation intensive and therefore requiring significant processing power. Therefore, hardware acceleration methods to increase the performance of such applications constitute an important area of study.

In this paper, we present an evaluation of one such method to process streaming data, namely multi-processor pipeline architecture. The hardware is based on a Multiple-Processor System on Chip (MPSoC), using a data encryption algorithm as a case study. The algorithm is partitioned on a coarse grained level and mapped on to an MPSoC with five processor cores in a pipeline, using specifically configured Xtensa LX3 cores. The system is then selectively optimized by strengthening and pruning the resources of each processor core. The optimized system is evaluated and compared against an optimal single-processor System on Chip (SoC) for the same application.

The multiple-processor pipeline system for data encryption algorithms used was observed to provide significant speed ups, up to 4.45 times that of the single-processor system, which is close to the ideal speed up from a five-stage pipeline.

*Index Terms*— Embedded Systems, Encryption, MPSoC, Multiple-processor Pipeline, Streaming Applications

## I. INTRODUCTION

### A. Streaming Applications

Simply saying, a streaming application is a program which keeps performing an operation on a stream of data. Such a program would keep reading data from a continuous input stream, perform the required operations on it, and output the processed data as a stream. The most obvious usage of streaming applications is in multimedia processing. Encoding or decoding of video, image and audio data are essentially of streaming nature. There the stream of input data may represent either a sequence of pixel information, video frames or an audio signal. For example, a simple JPEG encoding system would read blocks of pixel data of a raw image from the input stream, perform DCT transform, quantization, entropy encoding and Huffman encoding on the data and write the data out.

I. B. Nawinne, M. S. Wickramasinghe, R. G. Ragel and S. Radhakrishnan are with the Department of Computer Engineering, University of Peradeniya, Peradeniya 20400 Sri Lanka (phone: +94-81-567-0747; e-mail: isurunawinne@gmail.com).

Looking beyond the obvious, it can be observed that there are a number of applications which processes streaming data. Data packet processing in networking systems, data encryption and decryption, recording sensory data, mathematical processing systems, bioinformatics and many other scenarios could be identified to have a streaming nature. For devices that perform such processing on streaming data, factors such as performance, chip area and power consumption are obviously of critical importance. In the context of this project, the focus will be on the design of processing systems for such streaming applications with regard to the above mentioned factors.

### B. MPSoC

A System on Chip (SoC) is an application specific embedded computing device, designed to perform a specific task. Multiple-processor System on Chip (MPSoC), as the name suggests, is a processing system with several processor cores integrated on to a single chip substrate, and therefore the multi-core variant of SoCs. A typical MPSoC would consist of a set of processor cores, shared and dedicated memory blocks, interconnections, peripheral devices and interfaces. The MPSoC architectures scrutinized in this study are known as multiple-processor pipelines, and are intended for processing of streaming data.

### C. Multiple-processor Pipelines

The basic idea of having more than one processor core in a system is to achieve parallel processing. In order to make use of multiple processors for a single application program, the program should be partitioned and each partition should be mapped to a processor. One of the latest approaches in mapping a streaming application on to a multiple-processor system is to implement the system as a pipelined multiple-processor system (Fig. 1).

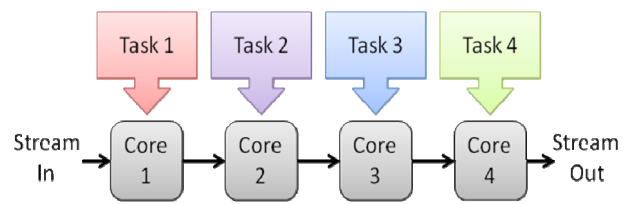

Fig. 1. A multi-processor pipeline: the cores are connected to each other in a sequential manner. Each partition of the program is mapped on to one of the cores.

There, each separate task (partition) of the streaming application program sits in a pipeline of processors. The primary idea of devising such a system is to improve the throughput by a significant amount. Ideally the throughput should be increased proportional to the number of pipelined processors used.

One of the biggest problems in instruction pipelines (inside a processor) is having to flush the pipeline completely, when a branch instruction breaks the flow, rendering the already partly executed instructions useless. Fortunately, there is hardly any such case in a task level pipeline for streaming applications and therefore multi-processor pipelines can be expected to provide a very high throughput.

The concept of heterogeneous multiprocessing comes into the picture when considering the efficiency of the processor pipeline. If all the processor cores in the pipeline are identical, it is likely that some processors might be highly utilized while others are underutilized, depending on the weights of the tasks running on each processor. Such situations may have bottlenecks created at those highly utilized points. Using tailor made processor cores for each task would get rid of this problem and avoid any such bottlenecks in the pipeline while increasing performance. There, the processor cores may be different from each other in various aspects, making the system heterogeneous.

In this paper, we present a multi-processor pipeline architecture customized for a product cipher algorithm. The hardware is based on an MPSoC. The algorithm is partitioned on a coarse grained level and mapped on to an MPSoC with five processor cores in a pipeline using a commercial tool for MPSoC configuration. The configuration details such as the architecture and the performance improvement due to the pipeline architecture are reported in this paper.

The rest of the paper is organized as follows: Section II lists and discusses related work in this area. Section III highlights the tools and technologies used and Section IV details our case study with the application considered, the product cipher. Details of the hardware system are reported in Section V and the results are presented in Section VI. The paper is concluded in Section VII.

## II. RELATED WORK

MPSoCs have been at the heart of many embedded devices for quite a while now. The concept of heterogeneous pipelined MPSoCs is fairly new in the research arena. Even though researchers have carried out numerous experiments regarding processor pipelining and heterogeneity of MPSoCs, few have looked into streaming applications of heterogeneous pipelined MPSoCs.

The researchers at Princeton have considered the advantages of using pipelined MPSoCs for streaming applications in general, identifying the improvements that could be made in multimedia and networking. They have looked mainly into multimedia applications, as in [1], such as audio/video compression, encrypting, etc. The researchers at UNSW have mainly targeted on the development of heterogeneous pipeline architectures and on the methods of their design space exploration. They have used a JPEG compression algorithm on various MPSoC architectures as a case study, as in [2] and [3], and carried out extensive research on its performance and the process of the design, as in [4].

An important aspect in designing multiple-processor systems or streaming applications is parallelizing the algorithm at the task level. Identifying various degrees of parallelizability possessed by different algorithms and efficiently mapping it into target architectures is not an easy task by any means. The sheer size of the design space for such a task is overwhelming. Thus, performing this automatically is obviously a much bigger challenge and a demanding area of research, attempted by many researchers. The Department of Computer Science at the University of Amsterdam along with the Leiden Embedded Research Centre at the Leiden University has developed framework named DAEDALUS, as in [5], which claims to perform this parallelization automatically with the aid of Kahn Process Networks (KPN) and efficiently explore the design space in order to obtain optimal results.

Taking a different approach about this, the Computer Science and Artificial Intelligence Laboratory at Massachusetts Institute of Technology has developed a programming language named StreaMIT, as in [6], specifically for implementing streaming algorithms. The structure of StreaMIT is designed such that the parallelization is obtained while implementing the algorithm. The application is developed targeting off-the-shelf processor architectures, and the designers claim that the application is optimized for the intended architecture.

When it comes to streaming applications, cryptography is a major subgroup in the research arena for a long time. Inherent streaming nature of most of the encryption technologies has made researchers to look deep into them and make use of them for experimenting on streaming processor systems irrespective of whether they are single processor systems or multi-processor systems. Even though researches on single processor systems for encryption are taken place throughout the world quite extensively and the outcomes of them are commercially available, researches on multiple-processor systems or MPSoCs are not so common.

Among the FPGA based multiple-processor solutions for cryptography, researches done by Department of Computer Engineering at Rochester Institute of Technology and Harris Corporation, RF Communications Division together takes a prominent place. Their FPGA-based, Multi-processor HW-SW System for Single-Chip Crypto Applications, as in [7], contains two isolated soft-core Nios II processors that share data through two crypto engines employs FPGA-based Single Chip Cryptographic techniques. These crypto engines are custom implementations of the Advanced Encryption Standard (AES).

The research, The Advanced Encryption Standard on an Asynchronous Shared-memory Multiprocessor, as in [8], done by Scott F. Smith of Department of Electrical and Computer Engineering at Boise State University can be considered as another prominent study where the Advanced Encryption Standard (AES) was utilized on a multiple-processor system architecture. His multiprocessor system uses an asynchronous bus system so that the scalability is increased. To achieve more scalability, Smith has used a globally-asynchronous locally-synchronous (GALS) design strategy over a globally clocked system. The processors used in this design are ARM922T processors.

From the many topics of interest within these researches, such as the various application scenarios and design process improvement, a primary and very significant area is performance optimization and the estimation of performance improvement gained by these architectures with respect to the chip area overhead.

The focus of this study is on the performance gains provided by a customized novel architecture, known as multiple-processor pipelines, for streaming applications.

### III. TOOLS & TECHNOLOGIES

Throughout this study, the host of tools provided by Tensilica's Xtensa environment, as in [9], is used. Tensilica Xtensa Xplorer IDE served as the cockpit throughout the whole process. A detailed description about the tools and technologies used are given below.

#### A. Tensilica Xtensa Xplorer

Tensilica Processor Developer's Toolkit is an integrated design environment used for comprehensive processor design, simulation, evaluation and optimization. The integrated development environment (IDE), Tensilica Xtensa Xplorer, is one of the major tools contained in Tensilica Processor Developer's Toolkit. This IDE with full graphical user interface was used as the centre of the whole designing, simulating and analyzing process of this research. Both single and multiple-processor modeling, simulation and analysis are performed on its graphical environment itself. In addition to these capabilities, it is fully integrated with Xtensa Software Developer's Toolkit to facilitate C application development, simulation, debugging, evaluation and optimization in the same development environment.

Xtensa Xplorer is also the gateway to the Xtensa Processor Generator, which generates the soft processor cores according to the designer's configurations. Once a processor configuration is created and finalized, it is automatically verified by the generator and the matching Xtensa Processor is built.

Xtensa Xplorer also provides facilities to estimate the actual chip area and power requirement of a configured core.

#### B. Xtensa LX3 Processor Cores

The Xtensa IDE facilitates the configuration of two types of Xtensa processor cores, namely Xtensa LX3 and Xtensa 8. In this study, Xtensa LX3 Processor Cores were utilized for the intended system designs. The ISAs of these cores are comprised of 24 bit instructions. They can even utilize 16 bit instructions when higher instruction densities are required. The sizes and other properties of data and instruction caches, RAMs and ROMs can also be specified.

#### C. Multiple-processor Systems

Two main approaches for modeling and simulating multiple-processor systems in Xtensa environment are Xtensa Modeling Protocol (XTMP) for modeling in C and Xtensa SystemC for modeling in SystemC, as in [10]. Both of these modeling tools provide APIs for Xtensa Instruction Set Simulator (ISS). In XTMP, simulations are described in standard C code.

In the Xtensa environment, multiple-processor systems are designed as multiple-processor subsystems. Xtensa Xplorer IDE automates the creation and development of these subsystems. During the designing phase of subsystems, the designer can configure many variables as per requirement. At the final stage applications to be run on each processor configuration are defined. After building all the projects and their shared projects, the subsystem is run, debugged or profiled as an XTMP simulation or an XTSC simulation.

#### D. Simulation

Xtensa Xplorer facilitates three simulation modes. The designer can choose either to run, debug or profile the designed system. Using the respective dialogs, the launch configurations used for these tasks can be created.

Xplorer IDE uses launch configurations as containers of details such as the binary, arguments, and simulation parameters. Xtensa supports two types of launch configurations for simulation purposes namely, Xtensa Single Core Launch and MP launch respectively for single and multi core systems.

### IV. CASE STUDY

To demonstrate the practical importance of this study, a set of encryption algorithms used in embedded applications were selected and a product cipher (combination of many encryption algorithms) was designed with those algorithms. Usually, embedded encryption algorithms are designed to be simple and to consume minimal processing power, thereby compromising the implied level of security. On the other hand, a product cipher would require higher processing power, while providing better security. Therefore a product cipher was designed, which would support the study with its reasonable complexity in computation and also provide sufficient ease for the partitioning process, which would otherwise consume a considerable amount of time. The selected embedded encryption algorithms for the product cipher are the following:

- *IDEA (International Data Encryption Algorithm)*
  IDEA, as in [11], was first proposed in 1991 to replace DES (Data Encryption Standard). It is a block cipher operating on 64-bit blocks of plain-text and uses a 128-bit key. The full IDEA algorithm runs 8.5 rounds, but the algorithm used here runs 8 rounds.

- *Skipjack*
  Skipjack, as in [12], was originally developed by the National Security Agency of USA in 1987 to be used in secure phones. It was declassified in 1998. Skipjack is a block cipher operating on 64-bit blocks of plain-text, uses an 80-bit key and runs 32 rounds.

- *Raiden*
  Raiden cipher, as in [13], was developed to replace the TEA algorithm which was proved easily breakable. It is a small block cipher operating on 64-bit blocks of plain-text using a 128-bit key. The version used here runs for 16 rounds.

IDEA and Skipjack algorithms require special processing to the key, before using it for encryption. Considering the computational complexity of the three algorithms, IDEA is the most complex, while Raiden is by far the least complex.

As depicted in Fig. 2, the product cipher created by combining these three algorithms also operates on 64-bit data blocks and uses a 128-bit key.

There, each 64-bit block of plain-text is first encrypted 20 times with the IDEA algorithm, then 24 times with the Skipjack algorithm and finally 20 times with the Raiden algorithm. The three ciphers are applied unevenly in order to make the effects of each cipher roughly balanced, which would help later when partitioning the algorithm. As the key for the Skipjack encryption in the product cipher, the first 80-bits of the 128-bit key are used.

A decryption algorithm was designed in parallel to this in order to verify the correctness of the product cipher, but only the product cipher was evaluated for performance.

## V. HARDWARE SYSTEMS

### A. Single-Processor System

First, a single-processor hardware platform was developed to run the product cipher benchmark. The system was built around an Xtensa LX3.0 processor core, and its configuration was calibrated to obtain the best performance for the application. This was done by adjusting the resources of the core and performing a number of test simulations. The features of the optimized single-processor system are as follows:

- Xtensa LX3.0 ISA
- Zero-overhead loop instructions
- Sign extend to 32-bit
- Normalize shift amount
- 16-bit / 32-bit multipliers
- 8KB 2-way associative instruction cache
- 8KB 2-way associative data cache

### B. Multiple-Processor Pipeline System

Next, the product cipher algorithm was divided into 5 partitions, to be implemented as a pipeline.

*Partition 1: Encrypting with IDEA, 10 times*
*Partition 2: Encrypting with IDEA, 10 times*
*Partition 3: Encrypting with Skipjack, 12 times*
*Partition 4: Encrypting with Skipjack, 12 times*
*Partition 5: Encrypting with Raiden, 20 times*

This way, the 64-bit plain-text blocks from the incoming stream could be passed through the five partitions sequentially, to ultimately achieve the exact same encryption as the original product cipher. Then, each of these partitions was mapped on to one of the five processor cores connected as a pipeline as illustrated in Figure 3.

The connections between the processor cores were made using shared memory. Every two adjacent cores in the pipeline share access to a buffer located in the shared memory, making a total of four buffers for the whole system. The shared memory was also used to make the encryption key available to all the cores.

Initially, identical processor cores were used for each of the five pipeline stages. Then, the system was simulated and the resource utilization was observed at each processor core in the pipeline. Based on the observations, excess resources were pruned at poorly utilized processors and extra resources were added to highly utilized processors which were creating bottlenecks in the pipeline. Actual chip area of the processor cores and core power requirements varied as these changes were made. This process was repeated a number of times to obtain reasonable optimization, without having to explore the whole design space, which is an extremely complex process by itself. Finally two systems were selected, one optimized to have minimal chip area and the other to have minimal power requirement.

The features of the processor cores of the optimized multiple-processor pipeline systems were as follows.

Common features of the cores

- Xtensa LX3.0 ISA
- Zero-overhead loop instructions
- Sign extend to 32-bit
- Normalize shift amount

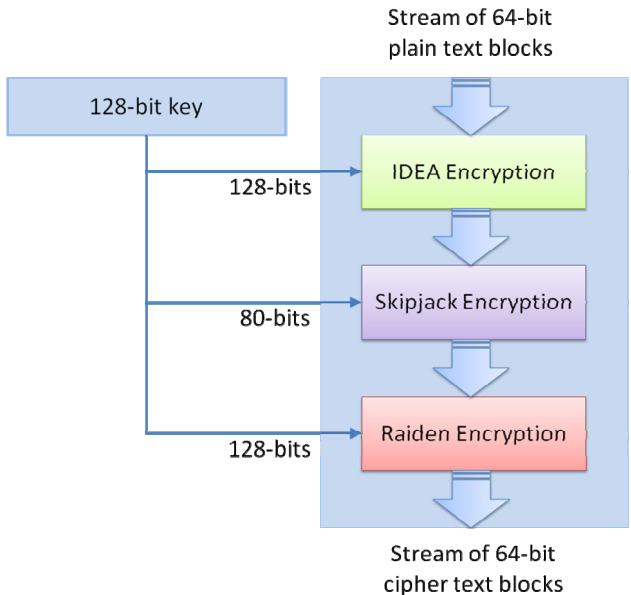

Fig. 2. The product cipher taken as the benchmark Streaming Application

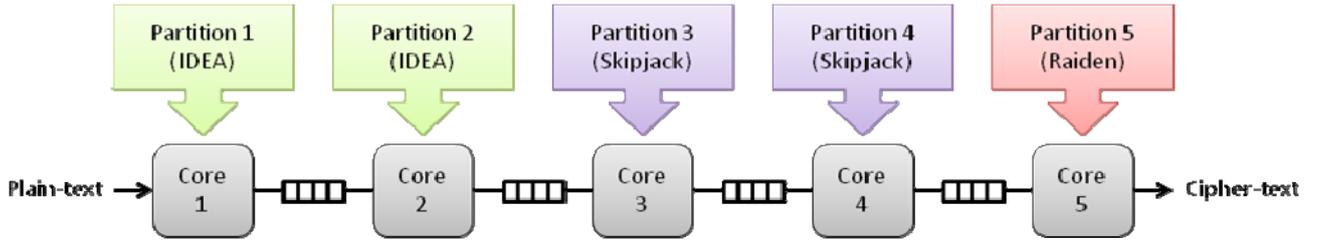

Fig. 3. Multiple-processor pipeline system for the product cipher: Five processor cores are connected as a pipeline, each core running one of the partitions from the product cipher. Connections between cores are made through shared memory.

Specific features for the power-focused system

*Core 1 and Core 2:*

- 4KB 2-way associative instruction cache
- 4KB 2-way associative data cache
- 16-bit/32-bit multipliers

*Core 3 and Core 4:*

- 8KB 2-way associative instruction cache
- 8KB 2-way associative data cache

*Core 5:*

- 4KB 2-way associative instruction cache
- 4KB 2-way associative data cache

Specific features for the area-focused system

*Core 1 and Core 2:*

- 4KB 2-way associative instruction cache
- 4KB 2-way associative data cache
- 16-bit/32-bit multipliers

*Core 3 and Core 4:*

- 8KB 2-way associative instruction cache
- 8KB 2-way associative data cache
- 16-bit/32-bit multipliers

*Core 5:*

- 4KB 2-way associative instruction cache
- 4KB 2-way associative data cache
- 16-bit/32-bit multipliers

Between these two optimal systems focusing on power consumption and chip area, the major difference is the use of 16 bit and 32 bit multipliers in all the cores of the area focusing system. The use of these functional units improves the performance of the systems while consuming more power and requiring less chip area.

Performances of these two multiple-processor pipeline systems were then evaluated and results were compared against those of the single processor system.

## VI. RESULTS

When evaluating the single-processor and multiple-processor pipeline systems for performance, all the cores were configured to have a clock speed of 200MHz. Both types of systems were simulated, inside the Xtensa Xplorer IDE using the Xtensa ISS (Instruction Set Simulator) and XTMP Simulations, with a sample plain-text stream as input. The running times, in microseconds, were calculated from the estimated number of total clock cycles as described below in (1).

$$Running\ Time = (Total\ Cycle\ Count) / 200 \quad (1)$$

For multiple-processor systems, the total cycle count for the whole system is the maximum of the cycle counts of all the cores in the system. For a single-processor system, the total cycle count is that of the core.

The recorded total clock cycle counts, for the systems simulated using the sample plain-text stream, are presented in Table I. The estimated running time, total chip area and total power are presented in Table II.

Here it can be seen that the multiple-processor pipeline systems consume more power and chip area as opposed to the single-processor system, but the time consumed to finish the processing is significantly less. Also obvious is the fact that the power-focusing multiple-processor pipeline consumes much less power than the area-focusing system, while requiring a higher chip area, due to the use of hardware resources in them as explained in the previous section.

The performance measure was defined and calculated from the running time as described below in (2).

$$Performance = 10^6 / (Running\ Time) \quad (2)$$

Next, the gains in performance achieved by the optimized multiple-processor pipeline systems, over the optimized single processor system, were calculated as in (3).

$$Performance\ Gain = \frac{Performance\ of\ Multiple\text{-}Processor\ Pipeline\ System}{Performance\ of\ Single\text{-}Processor\ System} \quad (3)$$

The performance gains were then evaluated with respect to the overheads of chip area and power of the optimized multiple-processor pipeline systems.

The performance gains (best values) achieved by the multiple-processor pipeline systems over the single-processor system are as follows.

- Overall Performance Gain = 4.448
- Performance Gain/Area Overhead = 0.996
- Performance Gain/Power Overhead = 1.664

TABLE I
RECORDED TOTAL CLOCK CYCLES FOR THE SIMULATED SYSTEMS

| Processor System | Total Clock Cycles |
|---|---|
| Single-processor system | 6,658,556 |
| Initial Multiple-processor pipeline | 2,292,090 |
| Multiple-processor pipeline (Power-focused) | 1,497,044 |
| Multiple-processor pipeline (Area-focused) | 1,496,844 |

TABLE II
ESTIMATED RUNNING TIMES, CHIP AREA AND POWER FOR THE SIMULATED SYSTEMS

| Processor System | Running Time (µs) | Total Chip Area (mm$^2$) | Total Power (mW) |
|---|---|---|---|
| Single-processor system | 33.293 | 0.45 | 55.74 |
| Initial Multiple processor pipeline | 11.460 | 3.65 | 62.6 |
| Multiple-processor pipeline (Power-focused) | 7.485 | 2.76 | 149.04 |
| Multiple-processor pipeline (Area-focused) | 7.484 | 2.01 | 278.7 |

## VII. CONCLUSION

The objective of this study was to explore the performance gains achievable by the processing architecture multiple-processor pipelines, for streaming applications. Using a product cipher encryption algorithm as a benchmark, partitioning it into five pipeline stages and implementing it as a multiple-processor pipeline, we have found that the performance can be improved by a factor of 4.448. Ideally the performance gain should be 5 for a five-stage pipeline, and the gain achieved is fairly close to that (89% of the ideal value). Conceptually, this value could be further improved by using better communication mechanism between processor cores, rather than using shared memory, thereby reducing the time spent on passing data between cores. The testing in this study was done using a sample stream containing 22 data blocks. Since the architecture of the hardware is a pipeline, better performance gains could be achieved for longer input streams. The hardware systems tested here were selectively optimized, in order to reduce the time spent on the optimization process. Performing a complete design space exploration can be expected to provide a better optimization for the system, with an added time cost.

The significance of this performance gain is that it is achieved with a comparatively less increase in the power consumption of the system. The gain in performance, relative to the increase (overhead) in power consumption is 1.664. This result is very important in the context of embedded applications, where the power consumption in a system is a critical matter. Also, the gain in performance observed, relative to the increase in chip area is 0.996, which is also a reasonable achievement. These aspects of the result imply that the performance improvement by using multiple-processor pipeline architecture is achieved with a well utilized increment of resources.

Generally, the performance improvement of a multiple-processor pipeline system lies in the parallelization of the streaming application in addition to the configuration of the hardware system. Therefore, it can be said that the benefits of using this architecture depends on the parallelizability of the application. If the number of partitions that the application could be divided in to is high, and also the complexity of the partitions are balanced, it is possible to achieve a very high gain in performance.

Ultimately, it can be concluded that by using multiple-processor pipeline systems, a substantial performance gain can be obtained for streaming applications.